\documentclass[letterpaper,twocolumn,english,aps,prb,groupedaddress,superscriptaddress,showpacs,amsfonts]{revtex4}
\usepackage{ae}
\usepackage{aecompl}
\usepackage[T1]{fontenc}
\usepackage[latin1]{inputenc}
\usepackage{float}
\usepackage{amsmath}
\usepackage{graphicx}
\usepackage{amssymb}

\makeatletter
\usepackage{babel}
\makeatother
\begin{document}

\title{Single Magnetic Atom on a Surface: Anisotropy Energy and Spin Density}

\author{Jing-Neng Yao}

\affiliation{Department of Electronics Engineering and Institute
of Electronics, National Chiao Tung University, Hsinchu, Taiwan}

\author{Chiung-Yuan Lin}

\affiliation{Department of Electronics Engineering and Institute
of Electronics, National Chiao Tung University, Hsinchu, Taiwan}

\date{\today}

\begin{abstract}
Studying single-atom magnetic anisotropy on surfaces enables the
exploration of the smallest magnetic storage bit that can be
built. In this work, magnetic anisotropy of a single rare-earth
atom on a surface is studied computationally for the first time.
The single adatom and its substrate surface are chosen to be
a Dysprosium (Dy) atom and a copper-nitrite surface, respectively,
where single transition-metal magnetic atoms on the same surface were
previously studied one atom at a time by scanning tunneling microscopes.
We propose unconventional $f$ and $d$ subshell symmetries
so that following the first-principles calculations,
simple pictorial analyses of the spin-density distribution can be performed for the first time,
independently for both a rare-earth atom Dy and a transition-metal Fe.
The magnetic anisotropy energy of Dy on the surface is calculated to be
a factor of five larger than the previous highest one,
reaching a record-high value of 31 meV.
\end{abstract}

\maketitle

\section{INTRODUCTION}
The desire of higher-density magnetic data storage is becoming
more emergent to conform to the explosive growth of today's
information industry. The continued downscaling of storage bits
will require significant enhancement of magnetic anisotropy energy
(MAE) per atom to overcome the fundamental, superparamagnetic
limit. Single-atom magnetic anisotropy on surfaces serves
as an important bottom-up approach of such enhancements.
A surface magnetic adatom no longer has the isotropic orbital
electronic structures of its free atom, and consequently its
spin-orbit couplings (SOC) vary in different spin orientations.
Such an anisotropy is not only technologically relevant to the
smallest magnetic storage bit that can be built, but also of
great scientific interest for its fascinating quantum effects
\cite{QM1,QM2,QM3}. Pioneering experiments
\cite{CoScience,FeScience,CoNaturePhysics,FeCuScience}
have been able to measure the MAE of single transition-metal atoms
and dimers on surfaces, followed by computational studies
\cite{CoScience,FeScience,Co-Pt,FeShick} that are consistent with
the measurements. Among those experiments, scanning tunneling
microscopes (STM) can even reach the precision such that magnetic
atoms can be built, manipulated, and measured one atom at a time
at desired atomic sites of a CuN surface
\cite{FeScience,CoNaturePhysics,FeCuScience}.
There is even a very recent STM study \cite{FechainScience} that can fabricate
a bistable atomic-scale antiferromagnet,
which enables a low-temperature demonstration of
dense nonvolatile storage of information.
However, the magnetic properties of such adatoms, e.g.~spin density,
have not yet been understood in simple, atomic-scale microscopic pictures.
Also, MAE of non-transition-metal atoms, the rare earth for example,
have not been explored, and are potential candidates with even higher
single-atom MAE so that one may build atomic-scale storage bits in
the future.

We start this study from first-principles calculations of a single
rare-earth atom Dysprosium (Dy) on a CuN surface. Analysis of
subshell quantum numbers, orbital shapes, and occupations of the
$4f$ orbitals leads to simple explanation of the calculated
spin-density distribution of the Dy atom. The success of such orbital
analysis is then duplicated to the previously studied single Fe
atom on the same surface \cite{FeScience,FeShick,FeZitko}, showing
that this simple picture works independently for both the
transition-metal $d$ and rare-earth $f$ orbitals. This work of Dy
on a CuN surface is also the first study of the MAE of a single
rare-earth atom on a surface, which is calculated by first
principles to be a factor of five larger than the previously
largest single-atom MAE, Co on the Pt surface \cite{Co-Pt}.

\section{COMPUTATIONAL APPROACHES}
We construct a supercell of 5-layer Cu(100) slabs,
place N and Dy atoms on the surface in the same way as the
density-functional-theory (DFT) studies of single Fe and Mn
adatoms \cite{FeScience,FeShick,MnDFT},
and then perform DFT calculations using the basis set of
all-electron full-potential linearized augmented plane wave \cite{win2k}.
A naive local-density or
generalized-gradient approximation makes the $4f$ orbitals be pinned
at the fermi level and causes a fractional $4f$ occupation,
inconsistent with the photoemission measurements of rare-earth
nitride bulks, and this problem was shown to be resolved
by performing a DFT+U calculations instead \cite{DyN}.
To find the correct $U$ and $J$ values for DFT+U,
we perform PBE+U calculations \cite{Anisimov,Liechtenstein,PBE96,win2k}
and inspect the $4f$ occupation number $n_{4f}$ of Dy on the CuN surface
with a series of $U$ and $J$.
When $J=0$, and $U$ ranges from $0$ to $9$ eV (noting that $U=0$ denotes PBE itself),
the occupation number stays at $n_{4f}=9.6$ regardless of the $U$ value used.
This means that with solely $U$ turned on, the unphysical fermi-level pinning of
one of the $f$ states remains the same as the naive PBE.
If $J$ is turned on to be a positive value $1.2$ eV at $U=9$ eV,
the $f$ state originally pinned at fermi-level lowers in energy,
and becomes fully occupied. However, the occupation number then becomes
$n_{4f}\approx10$, inconsistent with the typical $n_{4f}\approx9$ in
either a DFT+U calculation of a DyN bulk \cite{DyN} or
a recent quantum-chemistry, complete active space self-consistent field
(CASSCF) calculation of a ${\rm DyCl}_3$ molecule \cite{DyCl}.
If we first determine $U=6$ eV from a constraint PBE calculation \cite{Novak,win2k},
and are gradually turning on a negative $J$, we initially find that
the DFT self-consistent cycles do not converge before $J$ reaches $-0.8$ eV,
indicating unstable electronic structures. When $J$ reaches $-0.8$ eV,
\begin{figure}
\includegraphics[keepaspectratio,clip,width=7cm]{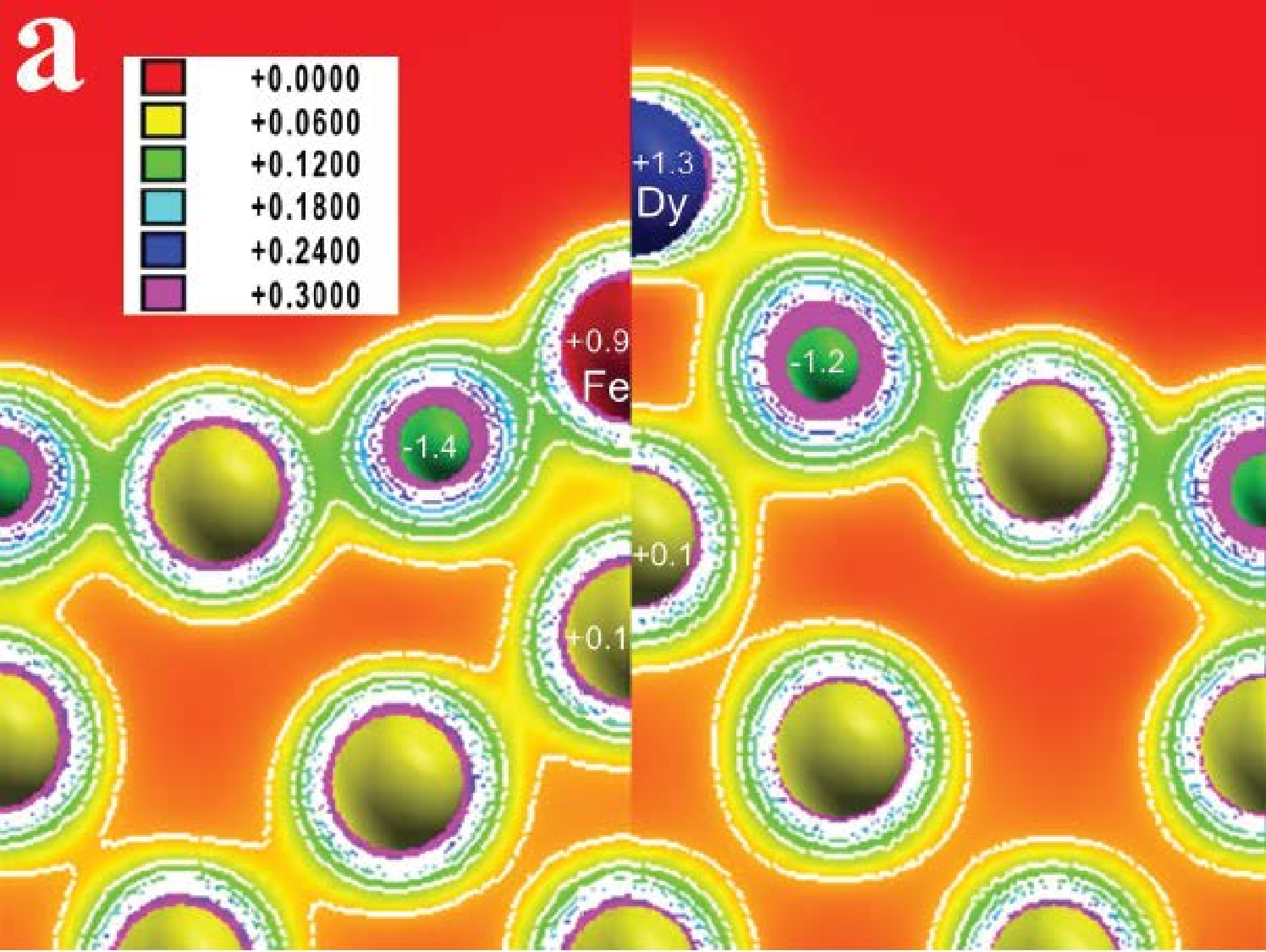}
\includegraphics[keepaspectratio,trim=0cm 10.5cm 3cm 0cm,clip,width=7cm]{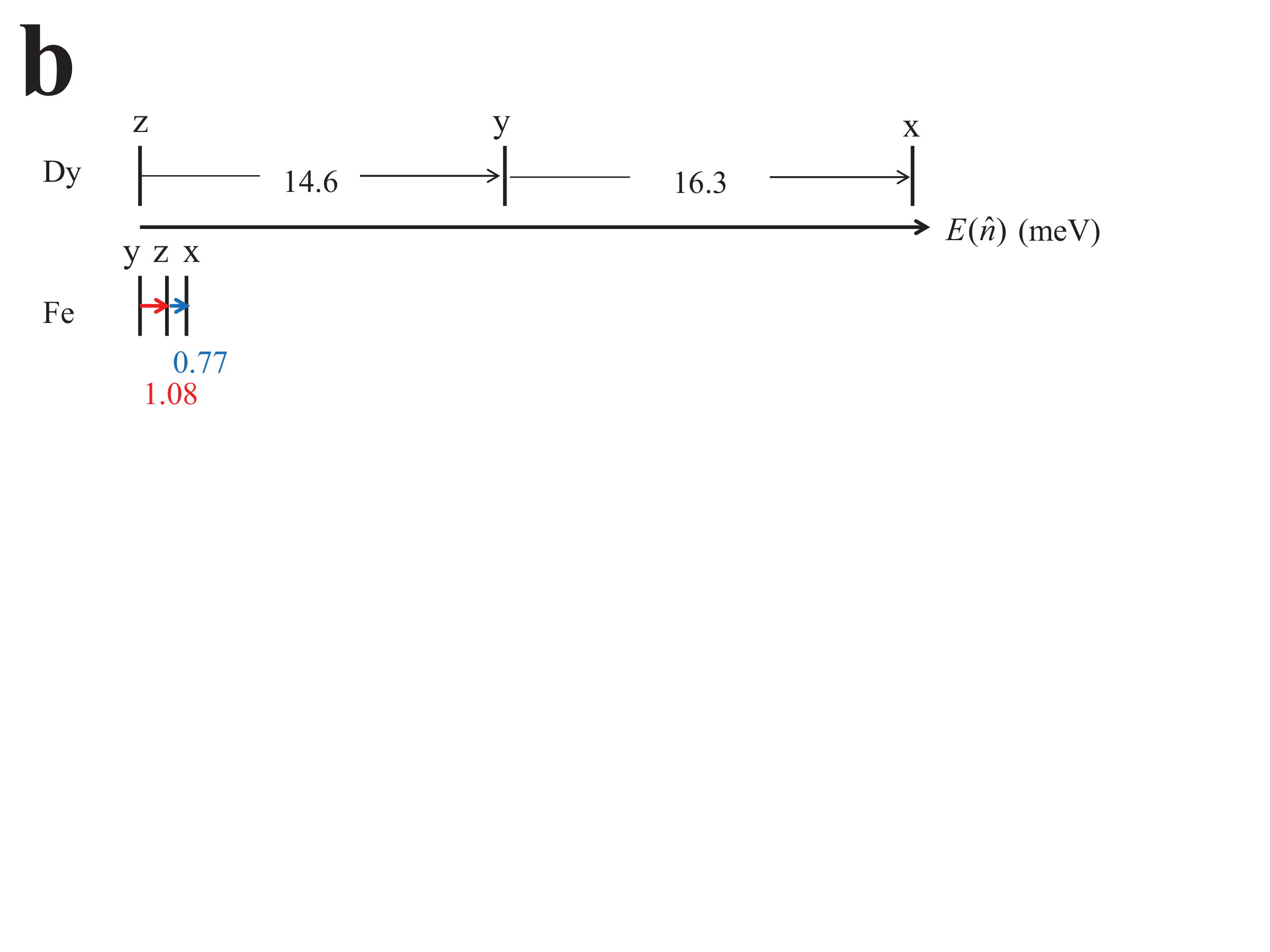}
\includegraphics[keepaspectratio,trim=0cm 0cm 0 0,clip,width=7cm]{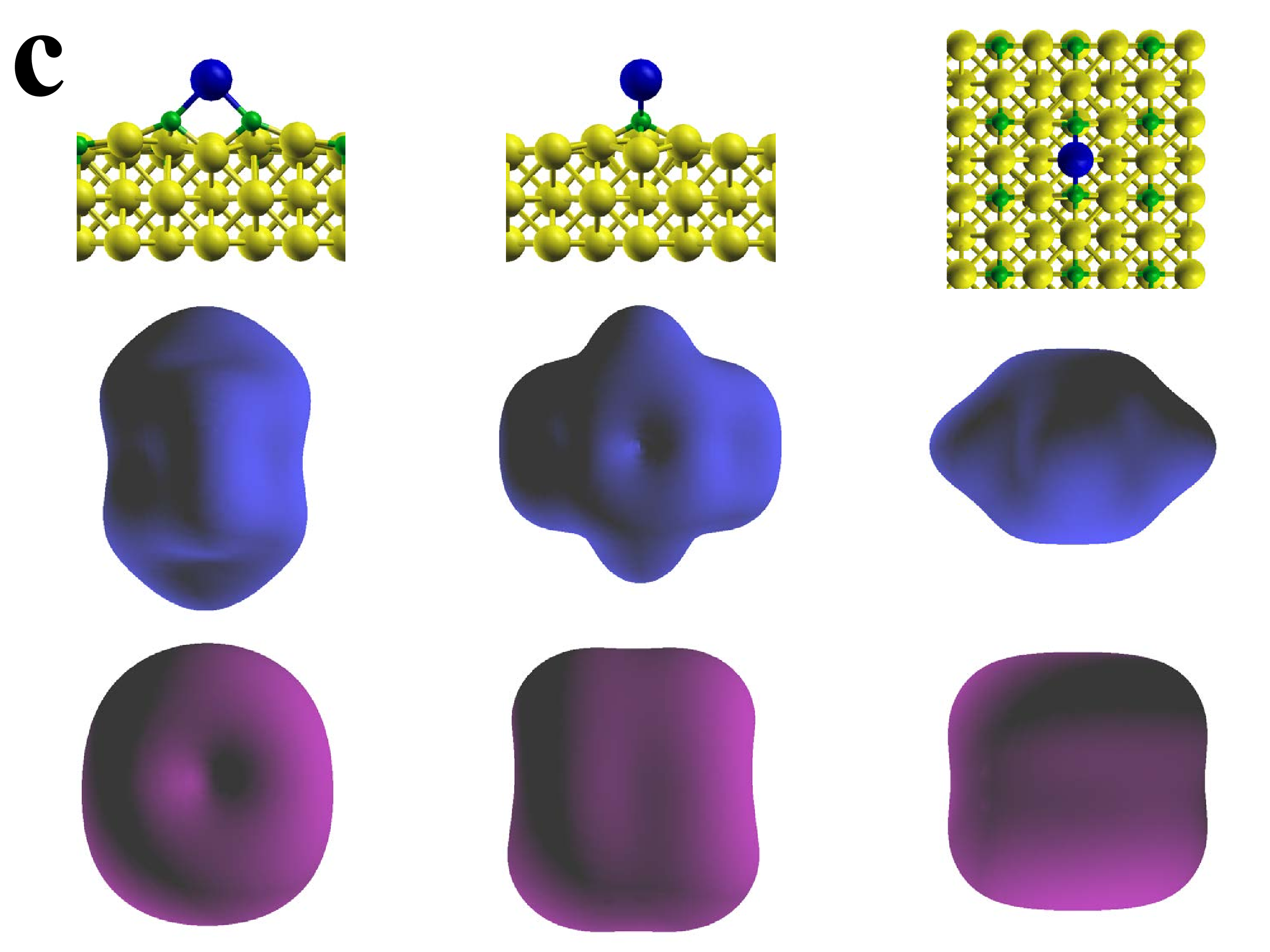}
\caption{({\bf a}) Electron density contour of a single Dy on the
CuN surface along the N-Dy-N raw and the out-of-plane direction,
in comparison with the Fe case \cite{FeScience}. The numbers
inside the circles indicate the net charge on selected atoms.
({\bf b}) Level diagrams in scale showing the calculated MAE of
Dy and Fe on the CuN surface, respectively.
The Fe case is done by Shick {\it et al.} \cite{FeShick} ({\bf c})
Calculated spin-density isosurfaces (second row, blue) of Dy on
CuN at the magnitude of $0.05e/a_0^3$ within a
$3\times3\times3$\AA$^3$ cube centered at the Dy nucleus, by
looking (left to right) along $x$, $y$, and $z$ directions. Each
stick-ball structure corresponds to one of the three directions.
The purple surfaces are the corresponding Fe spin density for
comparison.} \label{contour}
\end{figure}
the self-consistent cycles converge nicely, and give a reasonable
occupation number $n_{4f}=8.9$, consistent with the typical $4f^9$ configuration of Dy.
Therefore, for Dy on the CuN surface, we have determined that
$U=6$ eV and $J=-0.8$ eV yields a $4f^{9}$ configuration
that agrees with both the DFT+U calculation of a DyN bulk
\cite{DyN} and the CASSCF calculation of a ${\rm DyCl}_3$
molecule \cite{DyCl}. Atomic charges and spins are calculated by
Bader analysis \cite{Bader}.

\section{RESULTS}
As already pointed out in the previous Fe study \cite{FeScience},
when an adatom is deposited onto the Cu site of the surface,
it establishes polar covalent bonds with the nearest-neighbored N
atoms that replaces the original CuN binding network. The
calculated electron density of a Dy atom in the CuN surface is
shown in Fig.~\ref{contour}a, together with the previously
calculated Fe re-presented. As one can see, the Dy atom, sitting
even higher on top of the surface, attracts its neighboring N
atoms further out of the surface than the Fe case.
We have also calculated that Dy and its neighboring N
are $+1.3$ and $-1.2$ charged respectively.
Compared with the $+0.9$ and $-1.4$ charged Fe and
its neighboring N respectively, the Dy-N bond of the Dy system has
a polarity approximately the same as the Fe-N.

By pointing the Dy spin in the hollow, N-row, and out-of-plane
three symmetry directions (to be called $x$, $y$, and $z$
respectively) in our PBE+U total-energy calculations with SOC
included, we obtain the Dy MAE $E(\hat{n})$ of
$\hat{n}=\hat{e}_{x}$, $\hat{e}_{y}$, and $\hat{e}_{z}$, and
compare it with the Fe case, as shown in the level diagram in
Fig.~\ref{contour}b. One notices that in contrast to the Fe case,
the most-preferred magnetization axis of Dy is oriented in the
out-of-plane direction, while both atoms have their
least-preferred axis pointing in the hollow direction.
The MAE of Dy is basically one order of magnitude larger than that of Fe,
and is five times larger than the $6$meV MAE of Co on the Pt surface
\cite{Co-Pt}, the largest single-atom MAE reported previously.

The previously studied Fe on the surface has $13.5\%$ of spin
density extends into the surrounding atoms with the spreading
primarily along the N-row direction \cite{FeScience}. In contrast
to Fe, when calculating Dy on the same surface, we find that a net
spin of $S=2.9$ is localized at the Dy atom, and $S=2.9$ by
including the spin of all atoms, indicating that there is no spin
spreading. The $S=2.9$ of Dy includes two parts of contribution
from our analysis, $S=2.5$ from the localized electronic configuration
$(4f)^{9}$ (within the Dy muffin-tin sphere), and $S=0.4$ from the
delocalized $(6s5d)^{1}$
(out of the muffin-tin sphere but within the Bader basin),
where $6s5d$ denotes a hybridized molecular orbital \cite{6s5d}.
The significant reduction of spin spreading when
replacing Fe by Dy is obviously because the Dy $4f$ orbitals are
more localized than the Fe $3d$. Another interesting feature is
the shape of the spin density. When looking closely at the spin
isosurfaces of the Dy and Fe atoms along all three
crystal-symmetry directions in Fig.~\ref{contour}c, the shapes of Dy
and Fe are found to be approximately a hexagon and a square,
respectively, along either the N-row or out-of-plane direction,
while both atoms become more round-shape-like along the hollow.
This seemly mysterious observation of spin shapes will become clear
when we later look into the Fe $3d$ and Dy $4f$ orbitals in the CuN surface.
\begin{figure}
\includegraphics[keepaspectratio,trim= 0cm 0.3cm 2.5cm 0cm ,clip,width=7.5cm]{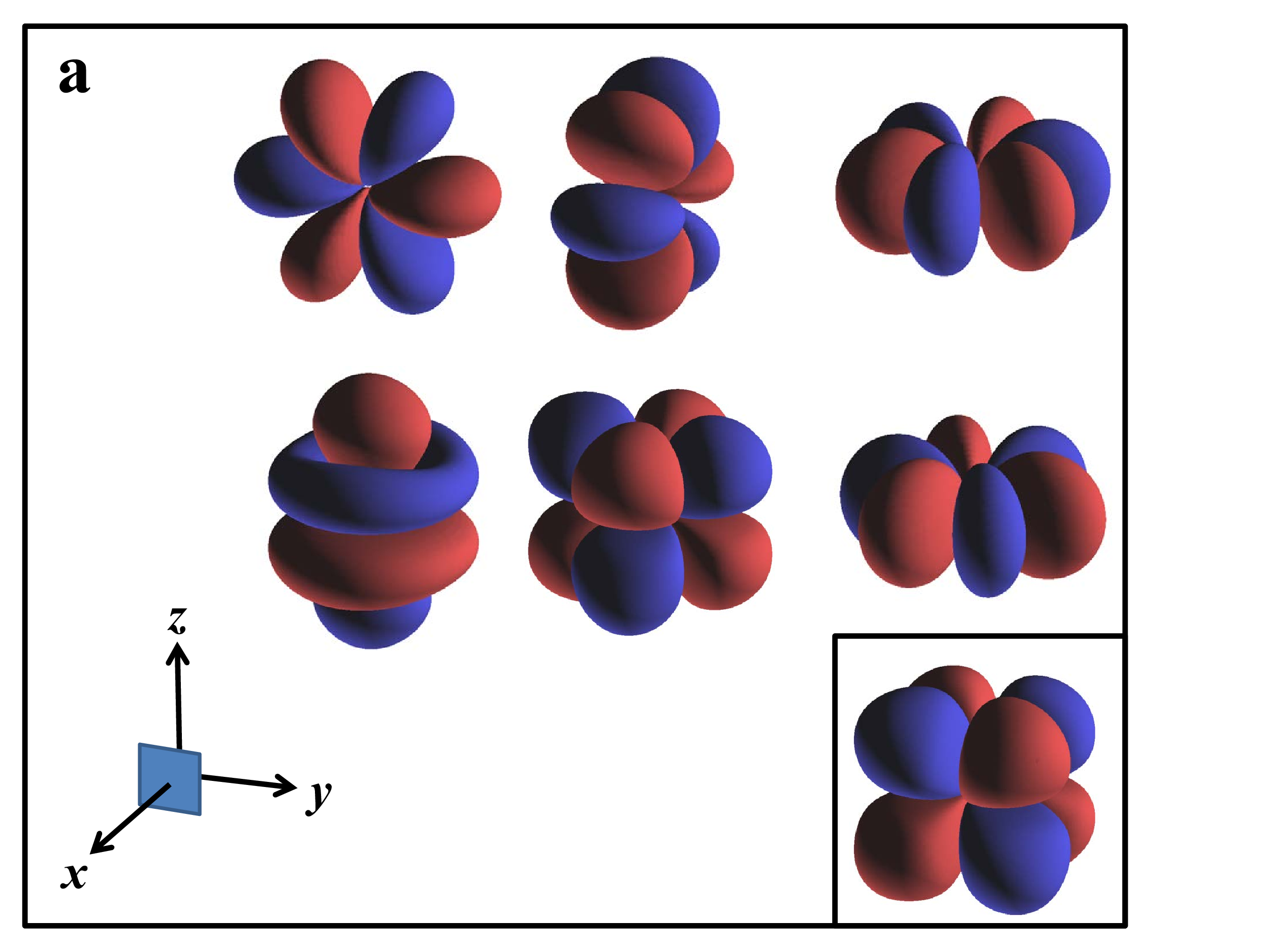}
\includegraphics[keepaspectratio,trim= 0cm 0.3cm 2.5cm 0cm ,clip,width=7.5cm]{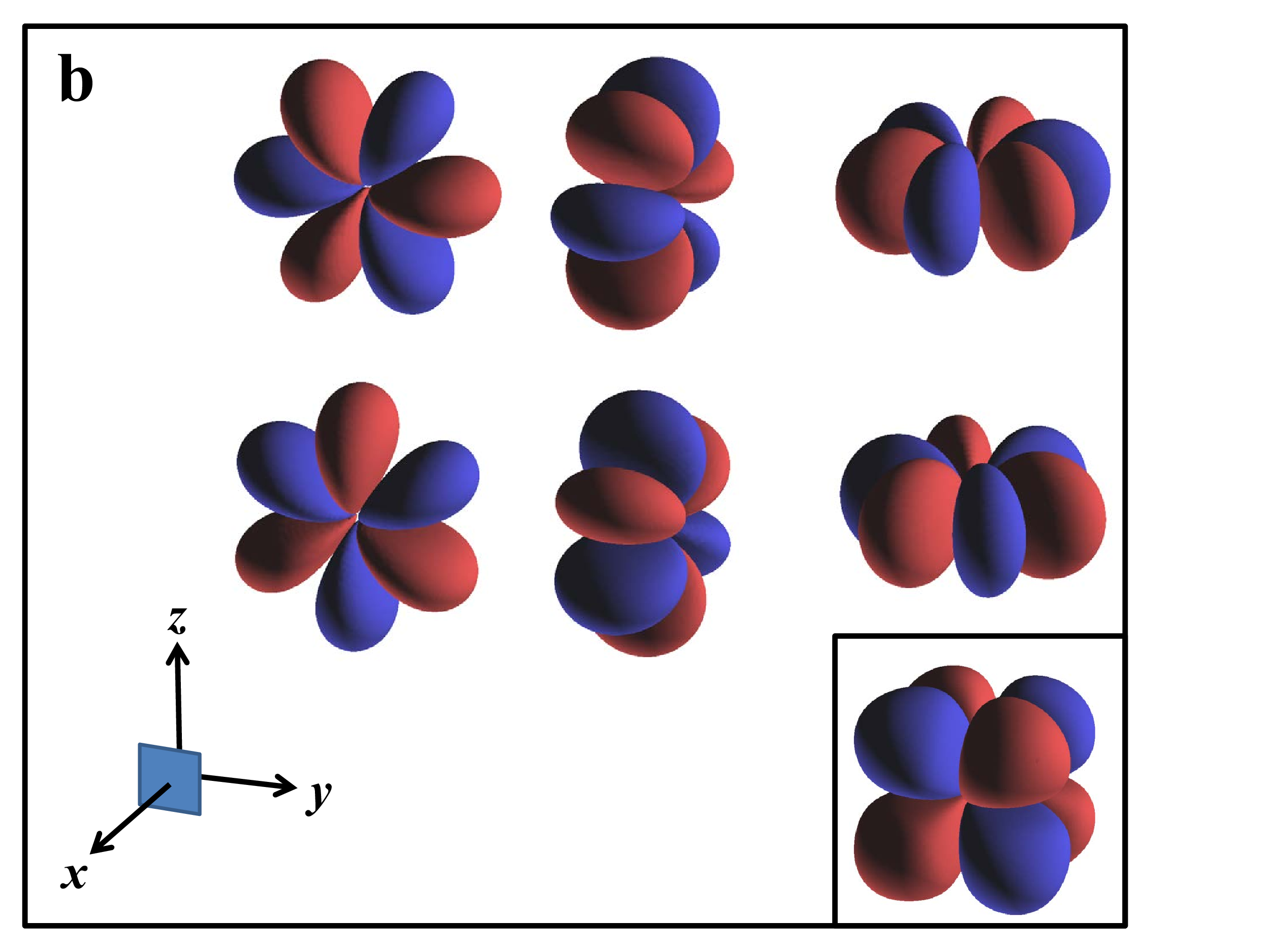}
\caption{\label{orbital}  Schematic plots of two different sets of $4f$ orbitals,
where blue (red) zones denote positive (negative) values.
({\bf a}) The general set \cite{orbitron}, from left to right:
(upper) $yz^2$, $z(z^2-3x^2)$, $x(x^2-3y^2)$,
(lower) $z^3$, $z(x^2-y^2)$, $y(3x^2-y^2)$.
The $xyz$ orbital is plotted in the inset at the right-lower corner.
({\bf b}) The $|L_{i}|$-fully-polarized set, from left to right:
(upper) $yz^2$, $z(z^2-3x^2)$, $x(x^2-3y^2)$,
(lower) $zy^2$, $xz^2$, $y(3x^2-y^2)$.
All six orbitals have a unique six-petal shape.
The only exceptional shape of the $xyz$ orbital is plotted
in the inset at the right-lower corner.}
\end{figure}

The $d$ orbitals in a crystal environment have the following
well-known subshell symmetries, $z^2$, $x^2-y^2$, $xy$, $yz$,
$zx$. The exotic, complicated $f$ orbitals, which are rarely shown
in literatures, have two commonly used sets of subshell
symmetries, the cubic and the general sets \cite{orbitron}.
As an example, we plot the $4f$ general set in Fig.~\ref{orbital}a.
We find there are four orbitals having a unique six-petal shape:
$yz^2$, $z(z^2-3x^2)$, $x(x^2-3y^2)$, and $y(3x^2-y^2)$, and are
different from the rest three. The six-petal shape suggests
that the hexagonal spin density is related to the $4f$ orbitals in some way.
However, the $z^3$ and $z(x^2-y^2)$ orbitals that do not have six-petal shapes
prevent us from establishing a relation between
the spin density and the general-set $4f$ orbitals.
Here we propose an unconventional set of $f$ orbitals such that six of
the orbitals have fully polarized angular momenta $|L_{i}|$,
as one will see its advantages in the orbital analysis of the spin shapes.
Five out of the seven orbitals, $xz^2$, $yz^2$, $y(3x^2-y^2)$,
$x(x^2-3y^2)$, and $xyz$ still belong to the general set. The rest two
are $z(z^2-3x^2)$ and $zy^2$, which are related to the general-set
orbitals $z^3$ and $z(x^2-y^2)$ simply by the following orthogonal
transformation
\begin{eqnarray}
&&\left(\begin{array}{cc}
    \left|x'({x'}^2-3{y'}^2)\right\rangle\\
    \left|x'{z'}^2\right\rangle
    \end{array}\right)
=\left(\begin{array}{cc}
    \left|z(z^2-3x^2)\right\rangle\\
    \left|zy^2\right\rangle
    \end{array}\right) \nonumber \\
    &=&\frac{1}{4}\left(\begin{array}{ll}
    \sqrt{10}&-\sqrt{6}\\
    -\sqrt{6}&-\sqrt{10}
    \end{array}\right)
    \left(\begin{array}{cc}
    \left|z^{3}\right\rangle\\
    \left|z(x^{2}-y^{2})\right\rangle
    \end{array}\right), \label{six-petal}
\end{eqnarray}
\begin{figure}
\includegraphics[keepaspectratio,trim= 0 17 0 0,clip,width=8cm]{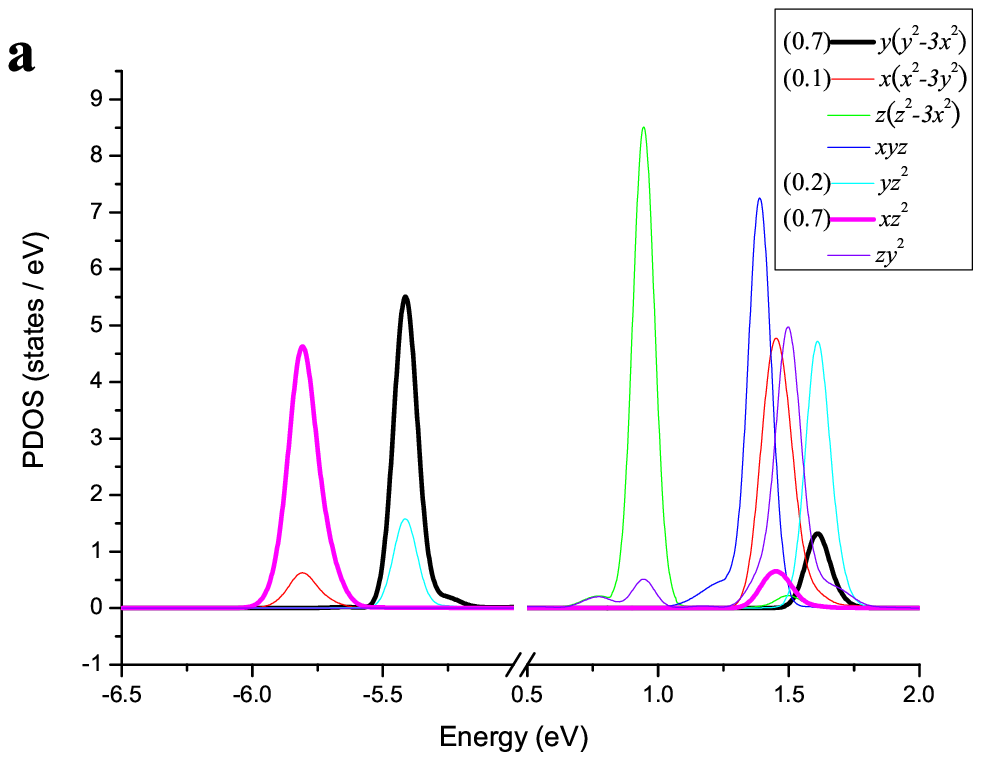}
\includegraphics[keepaspectratio,trim= 0 17 0 23,clip,width=8cm]{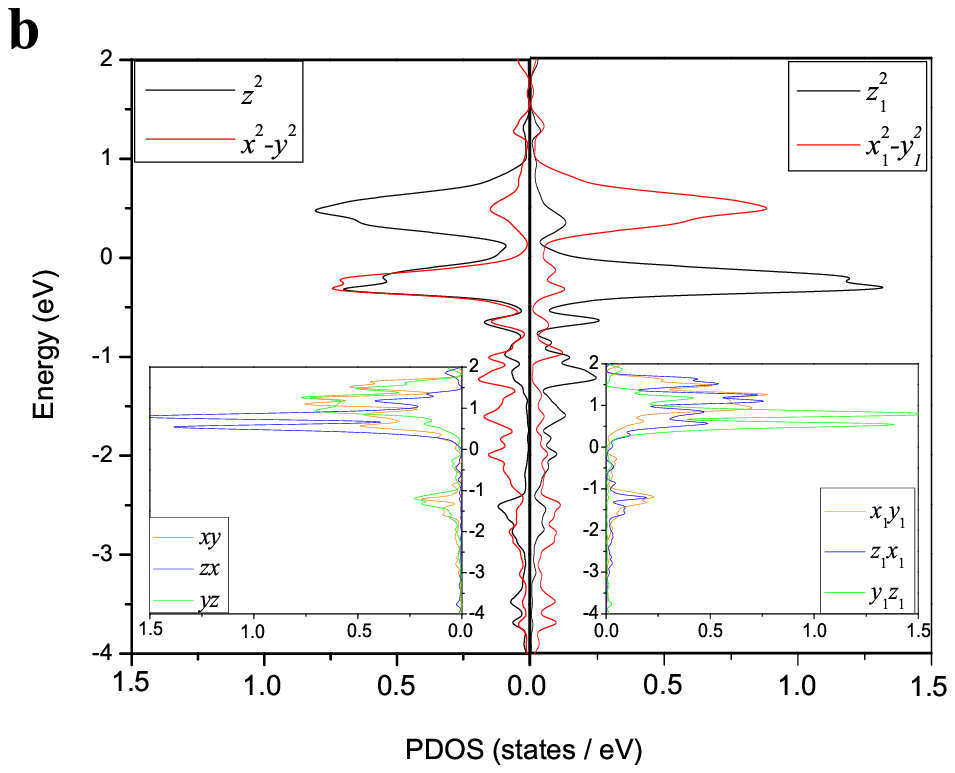}
\caption{\label{dos}
({\bf a}) The $4f$ minority-spin PDOS of Dy on the CuN surface without SOC.
The subshells follow the choice of  Fig.~\ref{orbital}b.
The nontrivial occupation are specified in parentheses.
({\bf b}) The $3d$ minority-spin PDOS of Fe on the
CuN surface. The left figure has the conventional quantization
axes with the hollow, N-row, and out-of-plane directions being
$x$, $y$, and $z$ axes, respectively
\cite{FeShick,FeZitko}. In contrast to the conventional set of
axes, the right one has unconventional axes
$(x_1,y_1,z_1)=(y,z,x)$, and is used to perform orbital analysis
in this work for its simple occupation numbers (either nearly
occupied or almost empty). The PDOS of three of the subshells
are plotted separately in each inset solely for the purpose that
all PDOS are better visualized.}
\end{figure}
where the primed coordinates are arranged as $(x',y',z')=(y,z,x)$.
It can be seen that $z(3y^2-z^2)$ and $zx^2$, which are
orthogonally transformed from the general-set $z^3$ and
$z(x^2-y^2)$, no longer belong to the general set in the unprimed
coordinates, but are actually the general-set orbitals
$x'({x'}^2-3{y'}^2)$ and $x'{z'}^2$ in the primed coordinates.
One can easily verify that each orbital of this set except for $xyz$,
with its quantization axis $\hat{n}$ properly chosen along one
coordinate axis $x_{i}$, has its angular momentum fully polarized,
that is, they are eigenstates of $|{\bf{L}}\cdot\hat{e}_{i}|=|L_{i}|$ with
eigenvalues $|m|=3=l$. Due to the reflection symmetry of these orbitals
in the $\pm\hat{e}_{i}$ directions, their angular-momentum polarization
has an sign ambiguity.
The full polarization of $|L_{i}|$ of these six $4f$ orbitals can be further
visualized clearly by looking at their orbital shapes. As one can
see in Fig.~\ref{orbital}b, the above mentioned six fully-polarized
$4f$ orbitals have an unique six-petal shape, and can be grouped
into three pairs. Each pair of orbitals have the same coordinate
axis as their central symmetry axis, and are related to each other
by exchanging their in-plane axes. The most obvious full
polarization can be realized for the $y(3x^2-y^2)$ and
$x(x^2-3y^2)$ orbitals, which are eigenstates of $|L_z|$ with
eigenvalues $|m|=3$, and have $z$ as their central axis.
Similarly, each of the rest four fully-polarized orbitals have
their central axes as their polarization directions. The only
exception, the $xyz$ orbital, has eight lobes pointing to the
corners of a cube.

Our calculations show that the Dy
$4f$ majority-spin states are all fully occupied and are very
low-lying, extremely atomic-like levels.
The orbital analysis of this work does not depend on the
details of the Dy $4f$ majority-spin states as long as all of them
are fully occupied, and their details are not presented here.
For minority-spin states, we plot their PDOS in Fig.~\ref{dos}a,
and find that $xz^2$ and $y(3x^2-y^2)$ are
mainly occupied, and $yz^2$ and $x(x^2-3y^2)$ slightly occupied.
The occupation numbers of the rest of the $4f$ minority-spin states are negligible.
The $5d6s$ electron mentioned in a previous paragraph
is mostly delocalized out of the Gd muffin-tin sphere,
and has a negligible PDOS within that sphere.
The Dy $4f$ orbitals are rather localized as indicated
from a previous paragraph that they have an approximate
$0.4$~\AA~radius.
It is a good approximation to think of these orbitals in the way
of electronic configurations of an atom, and therefore the
nontrivial occupation (neither fully occupied nor completely
empty) is determined for each orbital from the area under the
curve below the Fermi level. The PDOS thus implies an
approximately $4f^{9}$ configuration for the Dy atom itself.

The conventional quantization axes of the Fe $3d$ PDOS are
oriented in the way that the $x$ axis points along the hollow
direction, $y$ along the N row, and $z$ out of plane. Such PDOS
have been presented in Ref.~\cite{FeShick,FeZitko}, and show that
all the majority-spin states are fully occupied. We re-calculate
the PDOS and plot the minority spins in Fig.~\ref{dos}b. The
minority-spin PDOS show that $x^2-y^2$ is fully occupied, $z^2$
partially occupied, and the rest three minority-spin states
basically empty. The partially occupied spin-minority $z^2$
prohibits us from establishing a simple picture of either the
spin-density shape or the SOC. However, one may notice that the
$z^2$ and $x^2-y^2$ PDOS profiles depend nontrivially on the
choice (or interchange) of the coordinate axes, while the $xy$,
$yz$ and $zx$ (see the insets of Fig.~\ref{dos}b) have trivially
the same set of PDOS profiles with interchange of PDOS labels
corresponding to the interchange of the coordinate axes. Therefore
we search for $z^2$ and $x^2-y^2$ minority-spin PDOS of all three
possible assignments of coordinate axes, i.e., the hollow, N-row,
and out-of-plane directions are $x$, $y$, and $z$ axes,
respectively, and the cyclic permutations. In all three axis
assignments, all the majority-spin states are always fully
occupied, and the $xy$, $yz$ and $zx$ minority-spin always empty,
while the $z^2$ and $x^2-y^2$ minority-spin states have specially
simple occupations as shown in Fig.~\ref{dos}b under a particular
assignment of new axes: the hollow, N-row, and out-of-plane
directions are $z_1$, $x_1$, and $y_1$ axes, respectively. In this
new coordinate system, only the $z_1^2$ (or equivalently $x^2$ in
the old coordinates) orbital has paired spins, while the rest four
all unpaired.

\begin{figure}
\includegraphics[keepaspectratio,clip,width=8cm]{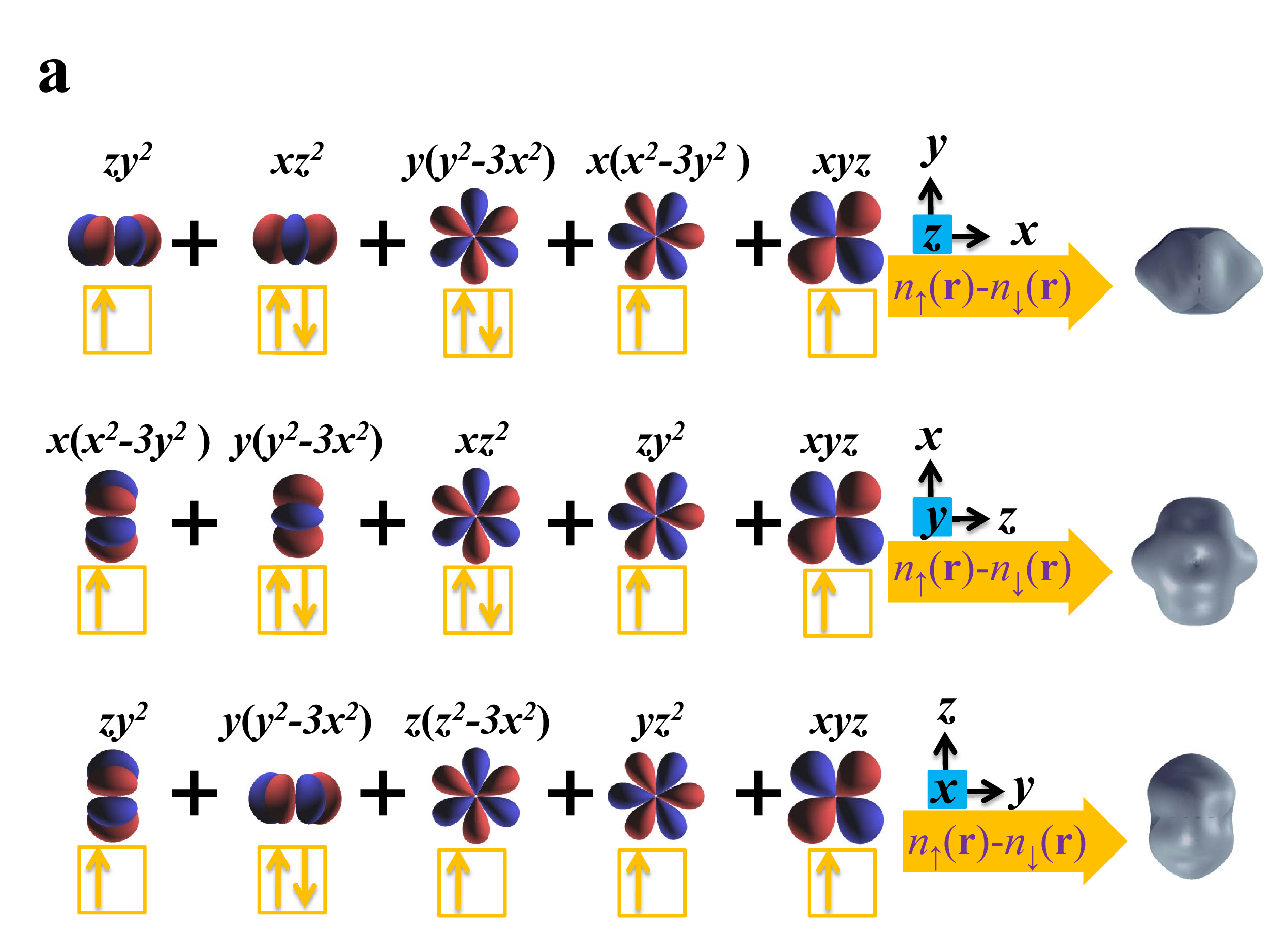}
\includegraphics[keepaspectratio,clip,width=8cm]{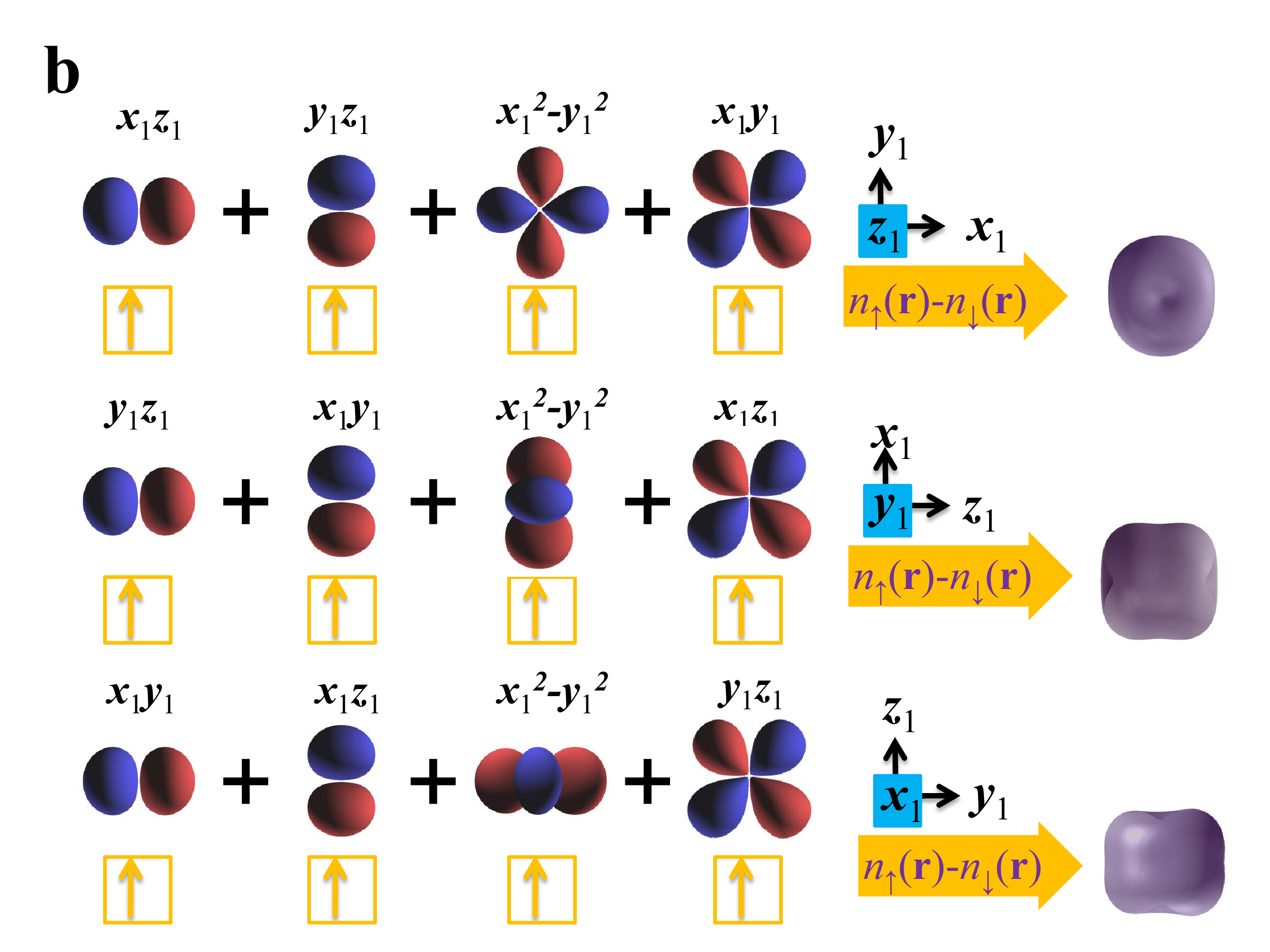}
\caption {Illustration of how analyses of localized orbitals explain
the shapes of the spin density of
({\bf a}) a Dy atom on the CuN surface, and
({\bf b}) Fe on the same surface.
In both cases, all spin up states are occupied, and
the spin-down occupation of each orbital is determined by the PDOS in Fig.~\ref{dos}.}
\label{schematic}
\end{figure}

\section{DISCUSSION}
With the orbital quantum numbers and occupation numbers
determined, we now try to explain the spin-density shape of a Dy
atom on the CuN surface. By starting with the top view, one
notices that the $x(x^2-3y^2)$ and $y(3x^2-y^2)$ orbitals both
have the six-petal shapes centered about the $z$-axis. As we have
identified from the PDOS analysis that $y(3x^2-y^2)$ has roughly
paired spins, while $x(x^2-3y^2)$ has roughly an unpaired spin.
The spin density shape along the $xy$ plane are therefore
dominated by the $x(x^2-3y^2)$ majority spin alone. Observing the
shape of the $x(x^2-3y^2)$ orbital, one then realizes that the
hexagonal shape of the Dy spin density from a top view in
Fig.~\ref{contour}c is essentially the consequence of the unpaired
$x(x^2-3y^2)$ orbital (see Fig.~\ref{schematic}a). Similarly, the
$z(z^2-3x^2)$ and $xz^2$ orbitals, both with a six-petal shape
centered about the $y$-axis, contribute to the hexagonal spin
density by looking along the $y$ direction. However, the $yz^2$
and $zy^2$ orbitals, centered about the $x$-axis, both have
unpaired spins. Their two six-petal shapes with $30^{\circ}$
relative to each other result in a slightly-round-shape
spin-density by looking along the $x$ direction.
We can even further understand the unique dent at the $y$ ends
of the spin isosurface by the following analysis.
The only unpaired orbital that carries a lobe in the $y$ direction
is $yz^2$, as can be seen in Fig.~\ref{schematic}a,
and all the rest have nodes along $y$.
On the other hand, both the other two axes receive lobe contribution from
two unpaired-orbitals, e.g., $x(x^2-3y^2)$ and $zy^2$ both contribute lobes along $x$.
Consequently, the spin isosurface shrinks in the $y$ direction due to
relatively less lobe contribution than the $x$ and $z$ directions.

The simple occupation picture of the adatom's localized orbitals
that explains the spin-density shape works not only for the rare-earth atom
Dy. As we are going to show below, the same picture also works for
the previously studied Fe adatom. When showing the Fe PDOS in both
the conventional quantization axes (hollow $x$, N-row $y$, and
out-of-plane $z$) and the unconventional ones (hollow $z_1$, N-row
$x_1$, and out-of-plane $y_1$) in Fig.~\ref{dos}b, we have
demonstrated that in the new coordinate system $(x_1,y_1,z_1)$,
only the $z_1^2$ orbital has paired spins,
while the rest four are all unpaired.
As seen from Fig.~\ref{schematic}b,
the square spin-density shapes of Fe in the CuN surface from the
top view and N-row side view are essentially consequences of the
spin-unpaired $z_1x_1$ and $y_1z_1$ orbitals, respectively,
while the round shape from the hollow-site side view is
the combination of the spin-unpaired $x_1^2-y_1^2$ and $x_1y_1$ orbitals.
One also notice that the spin isosurface has a dent only at the $z_{1}$ ends.
Similar to the way of explaining the Dy spin-isosurface dent,
such a dent is the consequence that both the $x_1$ and $y_1$ ends
have lobe contributions from the $x_1^2-y_1^2$ orbital,
while $z_1$ has no lobe contribution mainly due to the fact
that the $z_1^2$ orbital is spin-paired.

\section{CONCLUSION}
The STM moving-atom technique has demonstrated its capability of
building, manipulating, and measuring a single atomic spin in a
well-characterized environment. First-principles calculations
conclude that such an atomic spin forms a surface-embedded
molecular magnetic structure \cite{FeScience}, as well as,
reproduce the measured magnetic anisotropy axes \cite{FeShick}. As
an ongoing study of the STM-engineered adatoms, this work is the
first attempt to understand the spin density of a surface
magnetic atom in a simple, atomic-scale microscopic picture. This
is achieved by analyzing the occupations, shapes, and angular
momenta of its individual localized orbitals. These localized
orbitals include both the well-known $d$ orbitals of the
transition-metal atoms and the nontrivial $f$ of the rare-earth.
We determine an unconventional set of $4f$ subshell quantum numbers
and an unconventional set of $3d$ quantization axes
that can be used to analyze the shape of the spin density.
The spin-density shape of a magnetic adatom is explained by
simply counting the occupation of each individual subshell
and spin state of the atom's localized orbital.
These studies provide an important microscopic picture
of the atomic-scale origins of magnetization distribution.
We have also done the first computational study of
a single rare-earth atom on a surface.
The calculation predicts a record-high MAE of $31$ meV.
All these theoretical and computational realizations,
combined with continuing experimental innovations \cite{FechainScience},
may lead to further engineering of the single-atom magnetic anisotropy,
further exploration of the smallest magnetic storage bit,
and new fascinating applications of atomic-scale magnetism.

We thank A.~Heinrich, B.~Jones, and D.~Rugar for stimulating
discussions. We appreciate valuable technical support from
J.~L.~Li. We acknowledge financial supports from the Taiwan
National Science Foundation, the Taiwan Ministry of Education,
and the Taiwan National Center for Theoretical Sciences.
We also acknowledge the support of computing facilities provided by
the Taiwan National Center for High-performance Computing,


\begin{thebibliography}{99}
\bibitem{6s5d} The Dy $4f$, $5d$, and $6s$ have
radii of $0.4$, $1.4$, and $2.0$\AA, respectively. As a result,
the $4f$ contributes a much larger radial part $dV(r)/dr$ to the
SOC than $6s5d$. In fact, the $6s5d$ orbital is mainly distributed
out of the Dy muffin-tin sphere, and its SOC is not included in
this calculation.
\end{thebibliography}
\end{document}